\documentclass[
    reprint,
    superscriptaddress,
    preprintnumbers,
    nofootinbib,
    nobibnotes,
    aps,
    prd
    ]{revtex4-2}
\pdfoutput=1

\usepackage[english]{babel}
\usepackage{graphicx}
\usepackage[bookmarksopen, bookmarksnumbered, bookmarksopenlevel=2, hidelinks]{hyperref}

\hypersetup{
	pdftitle={Harmonic Analysis of the Instanton Prepotential},
	pdfauthor={Rafael \'Alvarez-Garc\'ia and Fabian Ruehle}
}

\usepackage{amsmath}
\usepackage{amssymb}
\usepackage{mathtools}
\usepackage{dsfont}

\usepackage{orcidlink}
\usepackage{soul}
\usepackage[all]{nowidow}
\usepackage{microtype}
\usepackage[capitalise]{cleveref}
\usepackage{enumitem}
\newcommand{\infdih}{\mathrm{I}_{2}(\infty)}
\newcommand{\CY}{\text{CY}_{3}}
\DeclareMathOperator{\rank}{rank}

\begin{document}

\title{Harmonic Analysis of the Instanton Prepotential}

\author{Rafael \'Alvarez-Garc\'ia\,\orcidlink{0000-0002-8927-2656}}
\email{ralvarezgarcia@fas.harvard.edu}
\affiliation{Department of Physics, Northeastern University, Boston, MA 02115, USA}
\affiliation{Jefferson Physical Laboratory, Harvard University, Cambridge, MA 02138, USA}

\author{Fabian Ruehle\,\orcidlink{0000-0002-8409-9823}}
\email{f.ruehle@northeastern.edu}
\affiliation{Department of Physics, Northeastern University, Boston, MA 02115, USA}
\affiliation{Department of Mathematics, Northeastern University, Boston, MA 02115, USA}
\affiliation{The NSF AI Institute for Artificial Intelligence and Fundamental Interactions}

\begin{abstract}
    Discrete symmetries of Calabi-Yau moduli spaces, generated by isomorphic flops, constrain the instanton expansion of the 4D $\mathcal{N}=2$ Type~IIA prepotential. We show that the Coxeter-invariant functions into which the prepotential organizes are eigenfunctions of a Laplace-Beltrami operator built from the Coxeter-invariant symmetric bilinear form on the moduli space. This means that the Gromov-Witten expansion can be interpreted as a superposition of waves propagating on the Coxeter quotient of the moduli space, and its resummation is the corresponding spectral decomposition. For the dihedral Coxeter groups, separation of variables in the eigenvalue equation explains from first principles why special modified Bessel functions, ordinary Bessel functions and Jacobi theta functions appear as the natural building blocks of the prepotential, depending on whether the Coxeter rotation acts hyperbolically, elliptically, or parabolically. The resulting spectral representations converge efficiently in the interior of the moduli space, complementing the standard large-volume instanton~expansion.
\end{abstract}

\maketitle

\section{Introduction}

Quantum corrections to effective field theories are often highly constrained by symmetry. In the context of string compactifications, topology-changing transitions of the internal geometry generate discrete symmetries of the moduli space that organize infinite families of quantum corrections into structured functions. In this letter, we show that this organization has a precise geometric origin: the instanton expansion of the low-energy effective theory can be interpreted as a superposition of waves propagating on the moduli space, and its natural resummation is the spectral decomposition of those waves into eigenmodes of a Laplace-Beltrami operator on the moduli space quotient. The resulting spectral representation converges efficiently in the moduli space interior, complementing the standard large-volume expansion, and explains the appearance of certain special functions, e.g., modified Bessel functions, ordinary Bessel functions and Jacobi theta functions, as the natural building blocks of the instanton prepotential.

More specifically, we study the effect of isomorphic flops (iso-flops) on the 4D $\mathcal{N}=2$ prepotential arising from Type~IIA string theory compactified on a \mbox{Calabi-Yau} threefold ($\CY$)~\cite{Bodner:1990zm}. Iso-flops are crepant small birational modifications connecting two chambers of the extended K\"{a}hler cone that correspond to diffeomorphic families of $\CY$ \cite{Aspinwall:1993yb,Aspinwall:1993nu,Greene:1996cy}. Each iso-flop wall behaves as a mirror in the K\"{a}hler moduli space, and a collection of them generates a Coxeter group $W$ acting on the K\"{a}hler moduli. As a consequence, the genus-zero Gopakumar-Vafa~(GV) invariants of non-flopping curve classes organize into \mbox{$W$-orbits} \cite{Lukas:2022crp,Candelas:2021lkc,Kuusela:2023vgi}, and the corresponding worldsheet instanton contributions to the prepotential must assemble into Coxeter-invariant functions $\psi_{ld}^{W}(T)$. Characterizing these functions is the central problem we solve.

Our main result is that each function $\psi_{ld}^{W}(T)$ satisfies a Helmholtz  equation for the Laplace-Beltrami operator $\Delta_\Sigma$ built from the symmetric bilinear form $\Sigma$ left invariant by the Coxeter action, i.e.,
\begin{equation}
    \Delta_{\Sigma} \psi_{ld}^{W}(T) = \lambda \psi_{ld}^{W}(T)\,,
\label{eq:helmholtz-main}
\end{equation}
for some $\lambda$ that depends on the case under study. This has an immediate physical interpretation: Each worldsheet instanton contribution $e^{2\pi il \langle d, T \rangle}$ can be interpreted as a plane wave in the moduli space, and the \mbox{Coxeter-invariant} function is a superposition of such waves over the group orbit. The wave equation~\eqref{eq:helmholtz-main} is satisfied by each plane wave individually and is inherited by the orbit sum through linearity. The resummation of $\psi_{ld}^{W}(T)$ into special functions is then nothing but the eigenfunction decomposition of this wave superposition in a basis that descends to the Coxeter quotient of the moduli space, with the specific special functions that arise determined by the geometry of the quotient.

Concretely, for the dihedral group $\mathrm{I}_2(m)$\,---\,which is the most prevalent case in the Kähler-favorable CICY classification \cite{Anderson:2017aux,ARPaper}\,---\,separation of variables in~\eqref{eq:helmholtz-main} reduces the radial equation to the modified Bessel, ordinary Bessel, or heat equation, depending on whether the Coxeter rotation acts hyperbolically, elliptically, or parabolically. This explains, from first principles, why these three families of special functions appear: they are the radial eigenmodes of $\Delta_\Sigma$ on the appropriate quotient geometry, selected by the physical boundary condition that instanton corrections decay exponentially at large volume. The two representations of $\psi_{ld}^W$, as a raw orbit sum and as a spectral eigenmode expansion, are dual descriptions; they are efficient at large volume and in the moduli space interior, respectively.

In the companion paper~\cite{ARPaper}, we present the complete CICY Coxeter Database, explicit formulas for all three dihedral cases, and the extension to general Coxeter groups via finite-state automata. The present letter focuses on the harmonic analysis perspective and its geometric origin, using the dihedral case as the explicit testing ground.
\section{Coxeter prepotentials}

A flop is a birational map between normal varieties that is an isomorphism in codimension-one and factors through a common singular model. In the context of smooth $\CY$, it contracts a collection of rational curves to points and replaces them by a different collection of rational curves. The locus in the extended K\"ahler cone at which these curves are completely shrunk defines a flop wall, separating two adjacent K\"ahler cone chambers. Whenever these two K\"ahler cone chambers correspond to diffeomorphic families of $\CY$, we say that we have an isomorphic flop, or iso-flop for short. The latter lead to integral involutions acting on the K\"ahler moduli or, dually, on the curve classes. A collection of these simple reflections generates a Coxeter group $W$ acting on the moduli space.

The genus-zero GV invariants of the non-flopping curves organize into orbits, i.e., $n_{d} = n_{wd}$ for all $w \in W$, where $d \in \mathcal{M}$ is a curve class in the Mori cone. As a consequence, the corresponding worldsheet instanton contributions to the superpotential must have the form \cite{Lukas:2022crp}
\begin{equation}
    \mathcal{F}_{\text{non-flop}} \coloneqq \sum_{d \in \mathcal{M}_{f}} n_{d} \sum_{l \in \mathbb{Z}_{>0}} \frac{1}{l^{3}} \psi_{ld}^{W}(T)\,,
\end{equation}
where $T$ represents the complexified K\"ahler moduli. The sum runs over the curve classes in a fundamental domain $\mathcal{M}_{f}$ for the action of the Coxeter group on the subcone containing the non-flopping curves. Here, $\psi_{ld}^{W}(T)$ stands for the $W$-invariant functions mentioned in the introduction. They are defined by summing over the $W$-orbit of a curve class $d \in \mathcal{M}_{f}$ as
\begin{equation}
    \psi_{ld}^{W}(T) \coloneqq \sum_{w \in W/\mathrm{Stab}_{W}(d)} e^{2\pi i l \langle wd, T \rangle}\,,
\end{equation}
where $\mathrm{Stab}_{W}(d)$ is the stabilizer of $d$ under the $W$-action. In favor of conciseness, we assume in the rest of the text that $d \in \mathrm{int}\left( \mathcal{M}_{f} \right)$, which implies $\mathrm{Stab}_{W}(d) = 1$, leaving stabilizer nuances for \cite{ARPaper}. Below, we focus on studying the properties of these $W$-invariant functions. Note that only a small number of flopping curve classes contribute to the instanton prepotential and that their interplay with the classical piece across the flop transition is well understood~\cite{Candelas:1993dm,Aspinwall:1993nu,Witten:1996qb}. This makes $\mathcal{F}_{\text{non-flop}}$ the most interesting part of the prepotential to study from the iso-flop perspective.
\section{Dihedral Coxeter prepotentials}

The dihedral group is the most abundant $\rank({W)} \geq 2$ group in the CICY Coxeter Database \cite{ARPaper}. Despite its simplicity\,---\,it is generated by just two iso-flops\,---\,its action on the moduli space leads to strong structural consequences for $\mathcal{F}_{\text{non-flop}}$. This is especially true for the infinite dihedral group subcase. Hence, explicitly studying the form of dihedral Coxeter prepotentials, building on the analysis initiated in \cite{Lukas:2022crp}, is a natural starting point that will allow us to gain intuition for the general case. Moreover, every edge in a general Coxeter diagram corresponds to a dihedral subgroup generated by a pair of simple reflections, meaning that all $\rank(W) \geq 2$ models contain the dihedral prepotential structure within them.

The dihedral group $\mathrm{I}_{2}(m) \cong D_{m} \cong \mathbb{Z}_{m} \rtimes \mathbb{Z}_{2}$ admits the presentation
\begin{equation}
    \mathrm{I}_{2}(m) = \langle r, s \mid s^{2} = 1, srs = r^{-1}, r^{m} = 1 \rangle\,,
\end{equation}
where $r$ and $s$ are the rotation and reflection generators, respectively. Consider the representation \mbox{$\rho: W \rightarrow \mathrm{GL}(V)$} with which $\mathrm{I}_{2}(m)$ acts on the curve classes, where \mbox{$h \coloneqq h^{1,1}(X)$} and $V \cong \mathbb{R}^{h} \supset \mathcal{M}$, and denote the matrices corresponding to the two generators by $\mathcal{Q} \coloneqq \rho(r)$ and $\mathcal{S} \coloneqq \rho(s)$. With this notation, the raw orbit sum can be written like
\begin{equation}
    \psi_{ld}^{\infdih}(T) = \sum_{k \in \mathbb{Z}} \left[ e^{2\pi i l \left\langle \mathcal{Q}^{k} d, T \right\rangle} + e^{2\pi i l \left\langle \mathcal{Q}^{k} \mathcal{S} d, T \right\rangle} \right]\,,
\label{eq:hyperbolic-infdih-raw-sum}
\end{equation}
for the infinite dihedral case, or
\begin{equation}
    \psi_{ld}^{\mathrm{I}_{2}(m)}(T) = \sum_{k=0}^{m-1} \left[ e^{2\pi i l \left\langle \mathcal{Q}^{k} d, T \right\rangle} + e^{2\pi i l \left\langle \mathcal{Q}^{k} \mathcal{S} d, T \right\rangle} \right]\,,
\label{eq:elliptic-dih-raw-sum}
\end{equation}
for the finite dihedral one. Depending on the properties of $\mathcal{Q}$, we can subdivide the study of the dihedral Coxeter prepotentials into the following subcases \cite{Lukas:2022crp,ARPaper}:
\begin{itemize}
    \item Elliptic, for which $\mathrm{ord}(\mathcal{Q}) = m < \infty$ and we obtain a representation of the finite dihedral group.

    \item Parabolic, for which $\mathcal{Q}$ is unipotent of index 3, i.e., $\mathcal{Q} = \mathds{1}_{h} + \mathcal{N}$ with $\mathcal{N}^{2} \neq 0$ and $\mathcal{N}^{3} = 0$. This implies $\mathrm{ord}(\mathcal{Q}) = \infty$, yielding a representation of the infinite dihedral group.

    \item Hyperbolic, for which $\mathcal{Q}$ is not unipotent and $\mathrm{ord}(\mathcal{Q}) = \infty$, resulting in a representation of the infinite dihedral group as well.
\end{itemize}
The elliptic and parabolic cases can only occur for $h > 2$, while the hyperbolic one is realized for $h \geq 2$ already.

For the sake of brevity, let us primarily focus on the hyperbolic case. Without loss of generality, assume that the two iso-flops are associated with the first two K\"ahler moduli and introduce the notation
\begin{equation}
\begin{aligned}
    T_{\parallel} &\coloneqq (T_{1}, T_{2})^{T}\,,
    &T_{\perp} &\coloneqq (T_{3}, \dotsc, T_{h})^{T}\,,\\
    d_{\parallel} &\coloneqq (d_{1}, d_{2})\,,
    &d_{\perp} &\coloneqq (d_{3}, \cdots, d_{h})^{T}\,.
\end{aligned}
\end{equation}
Denote the two-dimensional $d_{\parallel}$-subspace of $V$ by $V_{0}$ and the two-dimensional blocks of $\mathcal{Q}$ and $\mathcal{S}$ corresponding to the iso-flop directions by $Q = \left. \rho(r) \right|_{V_{0}}$ and $S = \left. \rho(s) \right|_{V_{0}}$, respectively. In the hyperbolic case, the full representation of $\infdih$ is a split extension of the two-dimensional one, meaning that we can find an extension shift vector $b(d_{\perp})$ untwisting the coordinates and leading to a block diagonal structure for the group action. We define
\begin{subequations}
\begin{align}
    y(d) &\coloneqq d_{\parallel} - b(d_{\perp})\,,\\
    y'(d) &\coloneqq Sy(d)\,,\\
    \mathcal{C}(d_{\perp},T) &\coloneqq \left\langle b(d_{\perp}), T_{\parallel} \right\rangle + \left\langle d_{\perp}, T_{\perp} \right\rangle\,.
\end{align}
\end{subequations}
The $\infdih$-invariant functions can then be expressed as
\begin{equation}
    \begin{split}
        \psi_{ld}^{\infdih}(T) = e^{2\pi i l \mathcal{C}(d_{\perp},T)} &\left[ S\left( l,y(d),T_{\parallel} \right)\right.\\
        &\quad \left.+ S\left( l,y'(d),T_{\parallel} \right) \right]\,,
    \end{split}
\label{eq:raw-orbit-sum-S-factors}%
\end{equation}
where
\begin{equation}
    S\left( l,z,T_{\parallel}\right) \coloneqq \sum_{k \in \mathbb{Z}} e^{2\pi i l \left\langle Q^{k} z, T_{\parallel} \right\rangle}\,.
\label{eq:hyperbolic-one-seed-orbit-sums}
\end{equation}
After Mellin-Poisson resummation \cite{Butzer1998FiniteMellin,Bardaro2018QuadratureMellin}, the above raw orbit sums can be recast into \cite{ARPaper}
\begin{align}
    S \big( l,z,T_{\parallel} \big) &= \frac{1}{u} \bigg[ K_{0}\big(\Omega\big(l,z,T_{\parallel}\big)\big)\\
    &\quad+ 2\sum_{m=1}^{\infty} K_{\frac{\pi i m}{u}}\big(\Omega\big(l,z,T_{\parallel}\big)\big) \cos\big(m\theta\big(z,T_{\parallel}\big)\big) \bigg]. \nonumber
\label{eq:hyperbolic-one-seed-orbit-sum-Bessel}
\end{align}
Here, $K_{\nu}$ is the modified Bessel function of the second kind and $u$ is defined through the eigenvalues of $Q$, which have the form $\lambda_{\pm} = e^{\pm 2u}$. We have also introduced
\begin{align}
    \Omega \left( l,z,T_{\parallel}\right) &\coloneqq 4\pi l\sqrt{-\mathcal{A}\left(z,T_{\parallel}\right) \mathcal{B}\left(z,T_{\parallel}\right)}\,,\\
    \theta \left( z,T_{\parallel} \right) &\coloneqq \frac{\pi}{2u} \log \left( \frac{\mathcal{B}\left(z,T_{\parallel}\right)}{\mathcal{A}\left(z,T_{\parallel}\right)} \right)\,,
\end{align}
with $\mathcal{A}\left(z,T_{\parallel}\right)$ and $\mathcal{B}\left(z,T_{\parallel}\right)$ defined via the decomposition
\begin{equation}
    \big\langle Q^{k} z, T_{\parallel} \big\rangle = \mathcal{A}\left(z,T_{\parallel}\right) e^{2ku} + \mathcal{B}\left(z,T_{\parallel}\right) e^{-2ku}\,.
\label{eq:QkzTparallel}
\end{equation}

The resummed expression \eqref{eq:hyperbolic-one-seed-orbit-sums} is more natural from the point of view of the Coxeter symmetry, since it is \mbox{Coxeter-invariant} term by term. Even though the elliptic case is a finite orbit sum, it can be recast into an analogous object involving ordinary Bessel functions of the first kind. The sense in which the resummed $\psi_{ld}^{\mathrm{I}_{2}(m)}(T)$ are more organic from the Coxeter point of view is clearer in the finite case: they reorganize the raw orbit sum in terms of functions of the generators of the algebra of polynomial invariants of the Coxeter group, realizing Glaeser’s theorem \cite{Glaeser1962FonctionsComposees,Barbancon2019FiniteReflection}. Finally, the parabolic case leads to an expression involving Jacobi theta functions, as was already noticed in \cite{Lukas:2022crp}.

Beyond its intrinsic structural interest, the resummed expression \eqref{eq:hyperbolic-one-seed-orbit-sums} also offers some practical advantages with respect to the raw orbit sum definition of $\psi_{ld}^{\mathrm{I}_{2}(m)}(T)$. The latter converges rapidly in the large volume region, where it localizes around a few leading instanton contributions while the rest of the worldsheet instantons are highly suppressed. The \mbox{Bessel-mode} decomposition acts as a spectral dual, meaning that this translates to a very flat distribution of terms and the need for many \mbox{Bessel-modes} to achieve equivalent convergence in the large volume region. In the deep interior of the moduli space, however, the opposite situation occurs. The raw orbit sum is broadly distributed among many exponential terms and its convergence rate is rather slow. The resummed expression, on the other hand, efficiently localizes around the first few \mbox{Bessel-modes}. This phenomenon is familiar from the Fourier analysis of waves, which motivates the next section.
\section{Harmonic analysis of the Gromov-Witten expansion}

The resummed expressions from the previous section and the special functions appearing in them have a natural geometric origin in terms of a \mbox{Laplace-Beltrami} operator defined in the \mbox{Coxeter-symmetric} moduli space. They are a decomposition of $\psi_{ld}^{W}(T)$ in a basis of its eigenfunctions. We proceed to summarize this harmonic analysis of the Gromov-Witten expansion, postponing a more complete exposition to \cite{ARPaper}.

An instanton contribution to the Gromov-Witten expansion has the form
\begin{equation}
    e^{2\pi i l \langle d, T \rangle} = e^{-2\pi l \langle d, t \rangle} e^{2\pi i l \langle d, b \rangle}\,.
\end{equation}
Here, the complex K\"ahler moduli $T^{i} = b^{i} + it^{i}$, with $i \in \{1, \dotsc, h\}$, are decomposed into the real K\"ahler moduli $\{ t^{i} \}_{i \in \{ 1, \dotsc, h \}}$ and their axionic superpartners $\{ b^{i} \}_{i \in \{ 1, \dotsc, h \}}$, appearing in the decomposition of the K\"ahler form $J = t^{i}D_{i} \in H^{2}(X,\mathbb{R})$ and the Kalb-Ramond field $B = b^{i}D_{i} \in H^{2}(X,\mathbb{R})/H^{2}(X,\mathbb{Z})$, respectively, where $\{ D_{i} \}_{i \in \{ 1,\dotsc, h \}}$ is a K\"ahler basis. Such a term can be heuristically interpreted as a plane wave in the moduli space, with the axions behaving as the angular variables on a real torus and the curve class $d \in \mathcal{M}$ acting as an integral wavevector via the natural pairing. The term dependent on the real K\"ahler moduli acts as an exponential damping, ensuring that the wave decays at large volume. Using the standard Laplace-Beltrami operator on the flat torus yields
\begin{equation}
    \Delta_{b} e^{2\pi i l \langle d, b \rangle} = -(2\pi)^{2} l^{2} |d|^{2} e^{2\pi i l \langle d, b \rangle}\,,
\end{equation}
from which we see that the Gromov-Witten expansion is, from this perspective, a decomposition of the instanton prepotential in terms of irreducible Fourier modes of the shift symmetry group of the axion torus. The $l$-fold cover terms corresponding to the class $ld$ act as higher harmonics of the basic wavevector $d$.

Let us now analyze the geometry appropriate for the hyperbolic case, for which we recall that $\big(V,\rho\big)$ is a split extension of $\big(V_{0},\left. \rho \right|_{V_{0}}\big)$. After untwisting by the extension shift vector $b(d_{\perp})$, the representation becomes block diagonal and its non-trivial piece corresponds to $\big(V_{0},\left. \rho \right|_{V_{0}}\big)$. The spectator directions simply contribute a flat additive piece to the Laplace-Beltrami operator, acting on the overall $e^{2\pi i l \mathcal{C}(d_{\perp},T)}$ factor in \eqref{eq:raw-orbit-sum-S-factors}, and we ignore them in what follows.

The two-dimensional block $\big(V_{0},\left. \rho \right|_{V_{0}}\big)$ of the Mori representation leaves a non-degenerate symmetric bilinear form $\Sigma$ with Lorentzian signature $(1,1)$ invariant. Dually, the same is true for the complexified K\"ahler cone. Define a complex quotient cylinder $C^{\mathbb{C}}/\langle Q \rangle$ with coordinates $(\Omega,\theta)$ and such that the $Q$-action corresponds to shifts of the angular variable $\theta$ by $2\pi$. From the metric
\begin{equation}
    ds^{2} = -d\Omega^{2} + \left( \frac{u}{\pi} \right)^{2} \Omega^{2}\, d\theta^{2}\,,
\label{eq:hyperbolic-complex-quotient-cylinder-metric}
\end{equation}
which respects the $Q$-action as isometries, we obtain the Laplace-Beltrami operator
\begin{equation}
    \Delta_{\Sigma} = \partial_{\Omega}^{2} + \frac{1}{\Omega} \partial_{\Omega} - \left( \frac{\pi}{u} \right)^{2} \frac{1}{\Omega^{2}} \partial_{\theta}^{2}\,.
\label{eq:hyperbolic-Laplace-Beltrami}
\end{equation}
Note that \eqref{eq:hyperbolic-complex-quotient-cylinder-metric} is a complexified version of the Rindler metric with Unruh temperature $T \propto u$. Define, at fixed curve class, the evaluation map
\begin{equation}
    \begin{split}
        E_{z,l} : \mathrm{int}\left( \hat{\pi}_{0} \left( \mathcal{K} \right) \right) &\longrightarrow C^{\mathbb{C}}\\
        T_{\parallel} &\longmapsto \left( \Omega \left( l,z,T_{\parallel} \right), \theta\left( z, T_{\parallel} \right) \right)\,,
    \end{split}
\end{equation}
where $\hat{\pi}_{0}$ is the projection to the $T_{\parallel}$-directions and $\mathrm{int}(X)$ denotes the interior of $X$. Consider the covering map $\pi: C^{\mathbb{C}} \rightarrow C^{\mathbb{C}}/\langle Q \rangle$ as well. The continuation of the evaluation map to the union of iso-flopped K\"ahler cone chambers is equivariant under the action of the rotation generator. Computing the pullback of the metric and the operator by both the evaluation map and its composition with the covering map, we obtain the \mbox{Laplace-Beltrami} operator $E_{z,l}^{*}\Delta_{\Sigma} = \big(\pi \circ E_{z,l}\big)^{*}\Delta_{\Sigma}$ with respect to the $T_{\parallel}$-variables, where we are abusing notation and denoting the differential operator on the complex quotient cylinder and on its covering space by the same symbol. One can then check that the Helmholtz equation
\begin{equation}
    E_{z,l}^{*}\Delta_{\Sigma} S \left( l,z,T_{\parallel} \right) = S \left( l,z,T_{\parallel} \right)
\end{equation}
is satisfied. In other words, \eqref{eq:hyperbolic-one-seed-orbit-sums} is an eigenfunction of $E_{z,l}^{*}\Delta_{\Sigma}$ that we have expanded in two different ways. Both the plane wave expansion and its resummation can be regarded as pullbacks of a superposition of eigenfunctions of the Laplace-Beltrami operator defined on the covering space of the complex quotient cylinder. The full sum is Coxeter-invariant and therefore descends to the complex quotient cylinder. However, each term in the plane wave expansion is not Coxeter-invariant individually; thus, they do not descend to the quotient and, in particular, cannot provide a decomposition in an eigenbasis of its \mbox{Laplace-Beltrami} operator. The resummation in terms of Bessel-modes, on the other hand, is expressed in a basis that is compatible with the quotient and is, therefore, a spectral decomposition pulled back from $C^{\mathbb{C}}/\langle Q \rangle$ rather than $C^{\mathbb{C}}$. It is in this sense that the resummed expressions for $\psi_{ld}^{W}(T)$ are more natural from the Coxeter-symmetry point of view, as was mentioned in the preceding section.

The origin of the special functions appearing in the prepotential is also clear from this perspective. Indeed, consider the eigenfunction decomposition
\begin{equation}
    S_{Q}(\Omega,\theta) = \sum_{m \in \mathbb{Z}} \phi_{m}(\Omega,\theta)\,,
\end{equation}
on $C^{\mathbb{C}}/\langle Q \rangle$ such that
\begin{equation}
    \Delta_{\Sigma} \phi_{m}(\Omega,\theta) = \phi_{m}(\Omega,\theta)\,,\quad \forall m \in \mathbb{Z}\,.
\end{equation}
Performing the separation of variables
\begin{equation}
    \phi_{m}(\Omega,\theta) = R_{m}(\Omega) e^{im\theta}\,,
\end{equation}
which is natural from the point of view of the isometries, the above equation reduces, after acting on the angular piece, to
\begin{equation}
    \Omega^{2}R''_{m}(\Omega) + \Omega R'_{m}(\Omega) - (\Omega^{2} + \nu_{m}^{2}) R_{m}(\Omega) = 0\,,
\end{equation}
with $\nu_{m} \coloneqq i \pi m/u$, which is the modified Bessel equation. This explains the special functions appearing in the resummation \eqref{eq:hyperbolic-one-seed-orbit-sum-Bessel} for the hyperbolic case.

Following the same steps for the elliptic case, for which the Mori representation is also a split extension of the two-dimensional block, we obtain the ordinary Bessel equation
\begin{equation}
    {\Omega'}^{2}R''_{r}(\Omega') + \Omega' R'_{r}(\Omega') + ({\Omega'}^{2} - \nu_{r}^{2}) R_{r}(\Omega') = 0\,,
\end{equation}
on the appropriate complex quotient geometry, where $\nu_{r} = mr$ with $r \in \mathbb{Z}_{>0}$ and $m \in \{ 2,3,4 \}$ for the $\mathrm{I}_{2}(m)$ groups found in the classification. For the parabolic case, the Mori representation is instead a non-split extension of the two-dimensional block and the non-trivial part of the problem lives in a three-dimensional subspace. Through similar means, we find that the Laplace-Beltrami operator in the corresponding complex quotient geometry reduces to the heat equation in one circular spatial dimension
\begin{equation}
    \partial_{\tau}\phi(\zeta_{3},\tau) = \frac{1}{4\pi i} \partial^{2}_{\zeta_{3}} \phi(\zeta_{3},\tau)\,,
\end{equation}
for which the heat kernel is the Jacobi theta function.

We expect the same phenomenon to occur for the general case: The Coxeter-invariant functions are defined as a raw orbit sum of worldsheet instanton contributions. Decomposing them into a superposition of eigenfunctions of the Laplace-Beltrami operator provides a resummed expression with complementary convergence properties to the original one. In \cite{ARPaper}, we comment on how to exploit the dihedral subgroups of a general $\rank(W) \geq 2$ Coxeter group to perform a partial resummation of $\psi_{ld}^{W}(T)$, since the dihedral Coxeter prepotential results can then be directly leveraged. A proper harmonic analysis of the remaining groups featured in the CICY Coxeter Database is left for future work.
\section{Conclusions}

We have shown that the Coxeter-invariant functions $\psi_{ld}^W(T)$ into which the non-flopping instanton contributions to the 4D  $\mathcal{N}=2$ Type IIA prepotential organize are eigenfunctions of a natural Laplace-Beltrami operator on the Coxeter quotient of the K\"ahler moduli space. The Gromov-Witten expansion, interpreted as a superposition of plane waves on the moduli space, can be rewritten as a spectral dual decomposition in terms of an eigenfunction basis of said operator. In the dihedral case, this explains the appearance of modified Bessel functions, ordinary Bessel functions and Jacobi theta functions, depending on whether the Coxeter rotation acts hyperbolically, elliptically or parabolically. The two representations are complementary: the raw orbit sum efficiently converges at large volume, while the spectral expansion does so in the interior of the moduli space.

Several directions follow naturally from this result. First, the physical content of the Helmholtz equation~\eqref{eq:helmholtz-main} deserves further investigation. Each worldsheet instanton contribution is a plane wave on the moduli space, and the wave equation selects the physical solutions through a large-volume boundary condition. Understanding which further constraints, for example on the prepotential, \mbox{attractor} flows, or distribution of vacua can be derived from this equation is an interesting open problem. Second, since higher-genus GV invariants also organize into \mbox{Coxeter} orbits~\cite{Kuusela:2023vgi}, the same spectral structure should arise for the higher-genus topological string amplitudes, with the genus-$g$ functions satisfying the same Helmholtz \mbox{equation} with genus-dependent coefficients. Third, the c-map connects the vector multiplet moduli space studied here to the hypermultiplet moduli space geometry at tree level; it would be natural to investigate whether \mbox{iso-flop} Coxeter symmetries impose analogous constraints on \mbox{D-instanton} corrections there. Finally, extending the classification of iso-flop Coxeter symmetries beyond complete intersections to the Kreuzer-Skarke database~\cite{Kreuzer:2000xy} would clarify how broadly the structures uncovered above apply across the landscape of Calabi-Yau compactifications.
\begin{acknowledgments}
    We thank Sarah Harrison, Vinicius Nevoa, Sanjay Raman, Cumrun Vafa and Kai Xu for useful discussions. The work of F.\,R. is supported by the NSF grants PHY-2210333 and PHY-2019786 (The NSF AI Institute for Artificial Intelligence and Fundamental Interactions). The work of R.\,A.-G. and F.\,R. is also supported by startup funding from Northeastern University.
\end{acknowledgments}

\bibliography{references}

\end{document}